# Logchain: Blockchain-assisted Log Storage

William Pourmajidi
Department of Computer Science, Ryerson University
Toronto, Canada
william.pourmajidi@ryerson.ca

Andriy Miranskyy
Department of Computer Science, Ryerson University
Toronto, Canada
avm@ryerson.ca

*Abstract*—During the normal operation of a Cloud solution, no one usually pays attention to the logs except technical department, which may periodically check them to ensure that the performance of the platform conforms to the Service Level Agreements. However, the moment the status of a component changes from acceptable to unacceptable, or a customer complains about accessibility or performance of a platform, the importance of logs increases significantly. Depending on the scope of the issue, all departments, including management, customer support, and even the actual customer, may turn to logs to find out what has happened, how it has happened, and who is responsible for the issue. The party at fault may be motivated to tamper the logs to hide their fault. Given the number of logs that are generated by the Cloud solutions, there are many tampering possibilities. While tamper detection solution can be used to detect any changes in the logs, we argue that critical nature of logs calls for immutability. In this work, we propose a blockchain-based log system, called Logchain, that collects the logs from different providers and avoids log tampering by sealing the logs cryptographically and adding them to a hierarchical ledger, hence, providing an immutable platform for log storage.

## I. INTRODUCTION

Logs are evidential documents [1]. That is, they contain the truth about the Quality of Service (QoS) that was delivered and can be used to draw conclusions that may affect the credibility of a service provider. Logs are also a key element in computer forensic investigations [2]. Log tampering has many forms and can be done by many different parties. Let us define tamper-motivation as the desire of one or more of the parties involved in a platform, infrastructure, or Cloud solution to access a critical log and to tamper this log by adding, removing, or manipulating a part of the log or the entire log. Below, we explore a few tamper-motivation situations that relate to each of the key participants in various types of Clouds.

In a private Cloud, where all stakeholders belong to the same company and (more or less) are trying to reach the same goal, a special type of tamper-motivation may exist. Imagine a financial company that has established a private Cloud. The management team has asked the IT department to establish a minute-by-minute backup of their data. Later, the company's primary storage is affected by a hardware failure. The IT department finds out that the per-minute backup has stopped working three days ago and had sent several alerts to the IT team, but the IT department has not checked these alerts. The IT department may be motivated to take advantage of having access to the logs, remove the alert messages, and show the management the tampered log, hence saving their jobs.

In a community Cloud, there are additional tamper-motivations. To survive, a community Cloud requires a clear definition of responsibilities, maintenance tasks, operational tasks, and control. Each partner is responsible for a subset of Cloud elements and together, all partner, ensure that the community Cloud remains operational and available. In case of an unfortunate incident, a party may be motivated to tamper the logs that show their fault, and, even worse, try to tamper the logs and fabricate a scenario in which another party becomes the main reason behind failure and, therefore, responsible for the caused damage.

In a public Cloud, many tamper-motivations exist. Clients using Infrastructure-as-a-Service do not have direct access to the bare-metal servers, core networks, and their related logs. Consider a scenario in which a public Cloud client deploys an application on an elastic Cloud environment. The client enables auto-scaling feature provided by the Cloud provider and defines a rule that when the memory usage exceeds 80%, the Cloud provider should allocate 20% extra memory space to the deployed application. Imagine that the client receives complaints related to the application performance from its users. The client suspects inadequate performance of the auto-scaling function as a root cause of the problem and asks the Cloud provider to send a detailed report of elastic memory allocation. The IT team of the Cloud provider checks their logs and finds out that the auto-scaling feature has worked intermittently, hence the performance issue. If the client finds out the truth, there may be a potential lawsuit on the horizon. Thus, the IT team may be motivated to tamper the log before sending it to the client.

To generalise the above-mentioned examples, in many incidents, there are parties who may be motivated to tamper the logs and in many cases, if the party succeeds, the incident, most likely, will forever remain a cold case. The same issue arises for the Cloud monitoring environment [3], [4]. A complete monitoring system requires full access to the Cloud resources and only Cloud providers have such level of access. Therefore, the majority of Cloud monitoring solutions are built by Cloud providers. While this is beneficial to the Cloud providers, it leaves Cloud clients with no option to verify the accuracy of provided monitoring details. One solution is to deploy the monitoring service on a trusted third-party platform that can be used by the Cloud client and the Cloud provider [3], [4]. However, this solution still requires a single trusted party, which is a constraint.

As shown above, the importance of a tamper-proof log system for Cloud solutions is significant. In other words, a traceable, verifiable, and immutable log system is required to establish trust among Cloud participants. Although a number of solutions for log tampering detection are available [5], [6], we argue that log tampering detection is not good enough and one should ensure that logs are tamper-proof, i.e., immutable.

The **goal** of this paper is to create a prototype of an immutable log system (called Logchain) using blockchain technology as a means to store and verify the collected logs. Our prototype constructs a Logchain-as-a-service (LCaaS) that receives logs or their hashes and stores them in an immutable hierarchical ledger; clients can use its API to interact with the solution and send, verify, and retrieve data from this immutable storage. The source code of the prototype can be accessed via [7].

## II. LITERATURE REVIEW

Digital forgery and tampering of digital artefacts and files long existed. Many solutions have been proposed to detect or prevent such undesired activities. Particular to files (which are the main form of storage for logs), various file verification techniques exist to ensure that the file at hand is not tampered. More than five decades ago, Peterson et al. described the use of cyclic codes to verify that a given array of bits is original or a change has happened [8]. Similar principles, known as checksum [9], [10], have been widely used in many areas to validate the integrity of files on various storage systems. One of the modern popular hashing techniques is a family of Secure Hash Algorithms (SHA) [11] which is used as a means to verify content, author or a property of a digital artefact. For example, source code management system git [12] generates SHA-1 [11] signature for a commit and uses it to trace the commit throughout the entire lifecycle of the source code [13]. However, the commits can be altered after the changes were committed, making them mutable. In addition to tools, many verification-as-a-service platforms offer integrity control for the uploaded data. Verification-or-integrity-as-a-service solutions, such as arXiv [14], offer a repository for electronic documents and ensure their integrity. The main drawback for these services is that one must trust the central authority (running the platform).

As for Cloud solutions, many of the previous methods are applicable. However, the complexity of Cloud environment (in particular, redundant systems and load balancers) and the scale of generated logs bring more challenges for the storage, access, and verification of the logs. Sharma [15] points out the complexity of mega-scale Cloud environment and suggest incorporation of various cryptographic algorithms and digital signature to achieve high integrity for storing critical information in the Cloud. Liu et al. [16] focus on the data storage integrity verification for Big Data in the areas of Cloud and IoT, stating that data integrity is critical for any computation-related system.

Bharath and Rajashree [17] suggest the use of a mediator, known as third-party auditor (TPA), which verifies the integrity of the data and sends the integrity report to the users. However, this still requires trust in a third-party or central authority. The problem of trust in the third-party can be alleviated by a properly implemented distributed ledger [18], [19], [20].

Current blockchain implementations of the distributed ledgers already have notary proof-of-existence services [18]; e.g., Poex.io [21], launched in 2013. Poex.io verifies the existence of a computer file at a specific time, by storing a timestamp and the SHA-256 [11] of the respective file in a block that will be eventually added to a blockchain. The service is anonymous: the files are not stored or transferred to the provider's servers. Since the digital signature of the file is permanently stored in a decentralised blockchain, the provider can verify the integrity and existence of such a file (at a point of submission to the blockchain) anytime in the future. Characteristics of cryptographic hash function [22] allow a provider to claim, with high certainty, that if the document had not existed at the time when its hash was added to the blockchain, it would have been very difficult to embed its hash in the blockchain after the fact. Additionally, embedding a particular hash and then adopting a future document to match the embedded hash is also almost impossible [22]. However, proof-of-existence solutions can not be used as a scalable log management systems, as they consider files individually, with no search function to locate the appropriate file or block. Moreover, Cloud solutions consist of thousands of hardware and software components, each of which generates large volume of logs [4]. The current solutions are not feasible for storing these volumes of logs, because the current public blockchains can handle limited number of concurrent transactions [18].

## III. LCaaS

LCaaS is a hierarchical blockchain framework, graphically shown in Figure 1. The figure depicts a two-level hierarchy, but the number of levels can be increased if a use-case requires it. Current blockchain consensus protocols require every node of the network to process every block of the blockchain, hence a major scalability limitation. We overcome this limitation by segmenting a portion of a blockchain and locking-it-down in a block of a higher-level blockchain, i.e., we create a two-level hierarchy of blockchains. Validating the integrity of a high-level block, confirms the integrity of all the blocks of the lower-level blockchain and leads to reduction of the number of operations needed to validate the chain.

We have built a prototype application that sits on top of a basic blockchain and converts it to a hierarchical ledger. Our primary goal is to bring scalability to blockchain for the situations in which the number of data items that need to be stored in a blockchain is large (e.g., operational logs of a cloud platform). At its current state, the prototype can load data from a log file and converts it to several blocks. Then the prototype will mine the blocks and puts them in a blockchain.

**Algorithm 1:** Generation of hash and nonce for a block. Our implementation instantiates Hasher using SHA-256.

**Input** : *block_index, timestamp, data, previous_hash*
**Output:** *current_hash, nonce*
1 *content* = concatenate(*index, timestamp, data, previous_hash*);
2 *content* = Hasher(*content*);   `// to speedup computing`
3 *nonce* = 0;
4 **repeat**
5    *nonce* = *nonce* + 1;
6    *current_hash* = Hasher( concatenate(*nonce, content*) );
7 **until** *prefix of current_hash = difficulty_target*;
8 **return** *current_hash, nonce*;

---

Finally, the prototype has the ability to convert the blockchain to circled blockchains and forms a hierarchical ledger. The current version of the prototype has not implemented the API component of the LCaaS.

*A. Key Elements*

*1) Blocks:* are atomic units of storage. Our implementation of blocks is similar to the existing blockchain solutions. A block contains the following variables: *nonce, index, timestamp, data, previous_hash,* and *current_hash*. *nonce* is an arbitrary random number that is used to generate[1] a specific *current_hash*. *index* is a unique sequential ID for each block. *timestamp* indicates the time when the block is created. *data* is a composite data type and contains information about logs. *current_hash* is generated by concatenating all of the above-indicated variables and adding the *current_hash* of the previous block, referred to as *previous_hash*. In other words, the *current_hash* of the $i$-th block becomes the *previous_hash* of block $i+1$.

One has to iterate through several values of *nonce*, to generate the *current_hash* for a given block that matches the defined *difficulty_target*. The target can be set during the initialisation of the LCaaS and may be adjusted later, if needed. The *difficulty_target* is often defined as the number of zeros that must appear at the beginning of the desired *current_hash*; the larger the number of zeros – the longer it will take (on average) to produce *current_hash* satisfying the *difficulty_target* requirement. Blocks are linked together based on a hash-binding relation. Formally, we show the creation of the *current_hash* in Algorithm 1.

*2) Blockchains:* blocks that are linked together will result in a blockchain. An $i$-th block in a blockchain relies on *current_hash* of block $i-1$ (as was discussed in Section III-A1); if data in an earlier block, say, block $m$ is tampered, the link among all the subsequent blocks, $m+1$ to $i$ will be broken and one will have to recompute *current_hash* (updating *nonce* values) of each block from $m$ to $i$. Mining, as a computationally expensive task, consists of taking the data in the Block, along with its *timestamp* and *previous_hash* and find a *nonce* that — when put together and hashed — results in a hash that matches the desired difficulty target of a blockchain. The difficulty target is often proposed as the number of required 0s at the beginning of the desired hash. Our implementation of blockchains and mining operations have the same characteristics of any other blockchain.

*3) Circled Blockchain (CB):* is a closed-loop blockchain that has a genesis and a terminal block (defined in Sections III-A4 and III-A5). The terminal block is the tail of a blockchain and indicates that the blockchain can not accept any more blocks. The terminal block converts a blockchain to the CB and makes it ready to be submitted to a new Superblock, defined in Section III-A6.

One needs to specify in advance the maximum number of blocks that can be appended to the CB or the maximum amount of time that CB stays 'open' until the terminal block is added to it, whichever comes first. These values would depend on the use-case. The goal is to create a CB with a reasonable number of blocks in it (denoted by $n_i$ for the $i$-th CB). If the frequency of log submission is high – a short window of time is preferred, otherwise, a larger window of time may be beneficial. The maximum amount of time should be fairly short to minimise the risk of tampering the whole CB: say, 24 hours or less.

*4) Genesis Blocks (GB):* is the first block of any blockchain. This block has predefined characteristics. Its *previous_hash* and *current_hash* are set to zero (as there are no prior blocks) and it has a null data element . Its primary purpose is to indicate the start of a new blockchain. We extended the genesis block definition, creating two different types of genesis blocks.

*Absolute Genesis Block (AGB)* is placed as the first block of the first blockchain. An AGB has the same characteristics as GB, with *previous_hash* and *current_hash* set to zero and data element set to null.

*Relative Genesis Block (RGB)* is placed at the beginning of every subsequent CB after the first CB. An RGB *current_hash* and *previous_hash* are set to the *current_hash* of the terminal block of the previous Superblock.

*5) Terminal Block (TB):* is similar to a genesis block, but it is added at the end of a blockchain to "close" it and produce a CB. The TB's data element has details about the CB that it has terminated. The elements are as follows. The *aggr_hash* is created by generating a hash (e.g., using SHA-256) of concatenated *current_hash* values of all blocks in that CB (AGB or RGB to the block prior to the terminal block). The data element may also store four optional values, namely *timestamp_from, timestamp_to, block_index_from,* and *block_index_to*. These optional values can be used by the search API to locate the required CB that contains the block or blocks that a user is looking for. Then, as with any other block, we produce a *current_hash* of the terminal block as per Algorithm 1.

---

[1]One may use different hashing functions. Currently, the most popular hash function in the blockchain community is SHA-256.

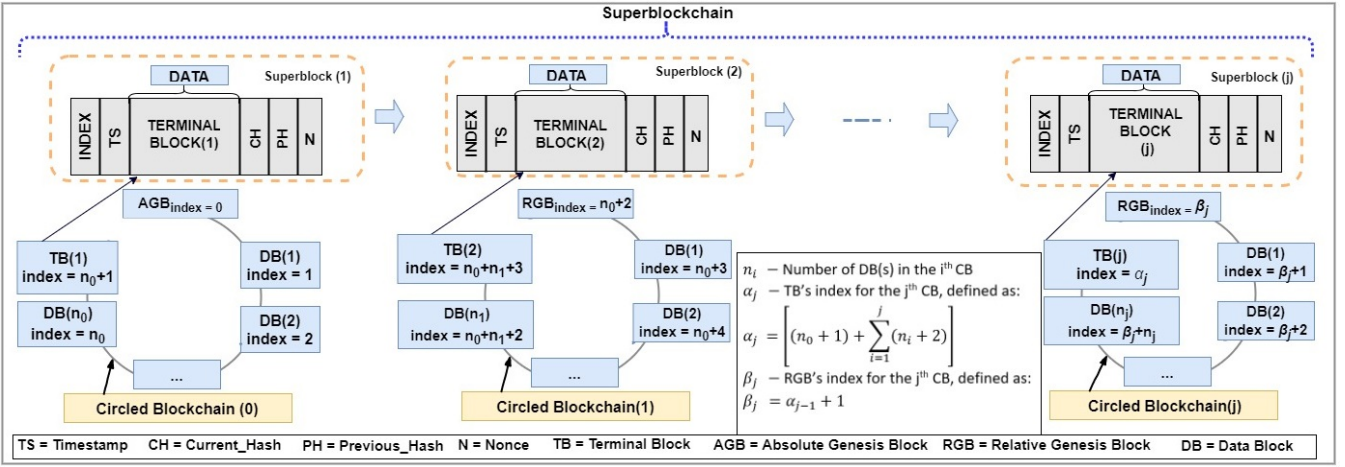

Fig. 1. Graphical representation of Superblockchain

*6) Superblock:* exhibits the features of a regular data block and has *nonce*, *index*, *timestamp*, *data*, *previous_hash*, and *current_hash*. The only differentiator is that its data element stores all of the field of a TB of a CB (*index*, *data*, etc.).

*7) Superblockchain:* is a blockchain consisting of Superblocks. The blocks are "chained": *current_hash* of a previous Superblock becomes *previous_hash* of the next one.

### B. API

The API, enabling users to interact with LCaaS, is as follows. There are two data submission functions: *submit_raw* and *submit_digest*. The former allows the client to submit the actual log file, the latter – just the file's digest (e.g., SHA-based digest computed using OpenSSL *dgst* [23]), thus, preserving the privacy of the log and reducing the amount of transmitted data. Both functions return, on success, timestamp and block_index of the transaction and, on failure, details of the error.

In addition to the file, the user may provide optional parameters that will be preserved in the blockchain, such as log file name, and starting and ending time stamps in this log file. These optional parameters may help to speed up the search for existing record in the blockchain, as discussed below.

For verification of an actual log file, one should use function *verify_raw*, for verification of the digest-based representation of the file – *verify_digest*. The functions would return the status of submission and number of blocks that matches the submitted data, if no block is found, the API will return zero. In case of an error, the API will return the failed status along with the error's description.

To improve the scalability of our solution, we introduced API function *verify_tb*. This function provides an assurance (in the cryptographic sense [22]) that a sequence of blocks was not tampered, as discussed in the beginning of Section III.

To improve accessibility of our solution, we introduced API function *search* that accepts a *block_index*, time interval based on block's *timestamp* values, or log records time intervals (using optional variables defined in Section III-A5 and III-A6) and returns the block elements as well as the TB elements associated with the block(s) found by *search*. The API functions may be implemented using REST and HTTP POST operation [24]; other methods of delivery are also acceptable.

### C. Practical Considerations

We built a prototype implementing core elements of the LCaaS [7]. However, for production implementations, we recommend building LCaaS on top of enterprise-grade blockchain services, such as IBM Blockchain [25] or Microsoft Blockchain [26]. For the private implementation, one can use one of the Hyperledger frameworks, such as Hyperledger Fabric [27], and build a hierarchical ledger on top of it. Furthermore, public blockchain services, such as Ethereum [28] can be used, but may not be financially feasible for a large number of logs. Essentially, one will need to create a single blockchain for Superblockchain and additional ones for each of the CBs.

## IV. ANALYSIS

LCaaS exhibits the following characteristics.

*Distributed Ledger* is shared between Cloud users and Cloud providers. Each participant has read-only and manageable access to some or all of the items in the ledger.

*Immutability:* hash of each block is created as per pseudo code shown in Algorithm 1. It incorporates the hash of a previous block; thus, any changes to the previous blocks would "break" the blockchain guarantying immutability.

*Cryptographically sealed:* nonces are used as a proof of work method to ensure that generated hashes meet the configured difficulty target. The hashes include all the elements in a block, including its timestamp, and nonce.

This hierarchical structure and its embedded recursive approach enhance the scalability, accessibility, and privacy of the hierarchical ledger compared to traditional blockchain platforms.

*Scalability Improvement:* relying on Superblocks, many Superblocks can be generated at the same time and then added to a Superblockchain at the same time. This will bring parallel processing feature for situations where multiple sources of data are generating data that needs to be put in the blockchain. For example, a platform may consist of twenty servers and each server can be associated with one Superblockchain.

*Accessibility Improvement:* API-based verification is added to the hierarchical ledger so users can submit raw data or digest values to check the consistency of their data.

*Privacy Improvement:* to improve privacy, an entire Superblockchain is reserved for a client to ensure that blockchains from different clients are not mingled. Furthermore, a user will only need to send the TB to the LCaaS to verify the integrity of the entire CB. Additionally, the option to store the hash value of data as opposed to real data, would bring additional confidentially to the clients.

## V. SUMMARY

The proposed LCaaS can act as a hierarchical ledger and a repository for all logs generated by Cloud solutions and can be accessed by all Cloud participants (namely, providers and users) to establish trust among them. Using verification services, a Cloud user can verify the logs provided by the Cloud provider against the records in the hierarchical ledger and finds out if the logs were tampered with or not.

In the future, we are planning to test LCaaS with existing blockchain solutions to find integration points that can be used to implement LCaaS on top of such solutions.

## ACKNOWLEDGEMENT

This research is funded in part by NSERC Discovery Grant No. RGPIN-2015-06075.